\documentclass{emulateapj}
\usepackage{color}
\slugcomment{Submitted to Astrophysical Journal Letters}

\lefthead{Romano-D\'{\i}az et al.}
\righthead{Constrained Cosmological Simulations of DM Halos}

\def\hmpc{h^{-1} \mathrm{Mpc}} 
 
\def\kmsmpc{\ {\rm km~s^{-1} Mpc^{-1}}}
 
\def\hmsun{h^{-1} M_\odot}
\def \etal {{\it et al.}}

\def \ie {{\it i.e.,} }
\def \eg{{\it e.g.,}}

\newcommand {\nbody} {$N$-body\ }
\newcommand {\rs} {$R_{\rm s}$}
\newcommand {\rvir} {$R_{\rm vir}\ $}

\begin{document}
\title{Constrained Cosmological Simulations of Dark Matter Halos}

\author{
Emilio Romano-Diaz\altaffilmark{1,2}, 
 Andreas Faltenbacher\altaffilmark{3}, 
 Daniel Jones\altaffilmark{2,4}, 
 Clayton Heller\altaffilmark{4},\\
 Yehuda Hoffman\altaffilmark{1},
 and Isaac Shlosman\altaffilmark{2}
}

\altaffiltext{1}{Racah Institute of Physics, Hebrew University,
Jerusalem 91904, Israel}
\altaffiltext{2}{Department of Physics \& Astronomy,
University of Kentucky, Lexington, KY 40506-0055, USA}
\altaffiltext{3}{Physics Department, University of California, Santa
Cruz, CA 95064, USA}
\altaffiltext{4}{Department of Physics,  Georgia Southern University, 
Statesboro, GA 30460, USA}

\begin{abstract}

The formation and structure of dark matter (DM) halos is studied by means
of constrained realizations of Gaussian fields using  
$N$-body simulations. A series of experiments of the
formation of a $10^{12}\hmsun$ halo is designed to study the dependence
of the density profile on its merging history. We confirm that 
the halo growth consists of violent and quiescent
phases, with the density well approximated by the Navarro-Frenk-White (NFW) 
profile during the latter phases. We find that (1) the NFW scale radius 
\rs\ stays constant during the quiescent phase and grows abruptly during the 
violent one. In contrast, the virial radius
grows linearly during the quiescent and abruptly during the violent
phases. (2) The central density stays unchanged during the quiescent
phase while dropping abruptly during the violent phase. 
(3) The value of \rs\ reflects the violent merging history of the halo,
and depends on the number of violent events and their fractional magnitudes,
independent of the time and order of these events. It does not
reflect the formation time of the halo.
(4) The fractional change in \rs\ is a nonlinear function of the
fractional absorbed kinetic energy within \rs\ in a violent event.
\end{abstract}


\section{Introduction}
\label{sec:intro}

The problem of the formation and structure of dark matter halos
constitutes one of the outstanding challenges of modern cosmogony and
structure formation models. The problem is easily formulated as what
is the outcome of the collapse and virialization of bound
perturbations in an otherwise homogenous and isotropic expanding
universe. This is further simplified by considering only collisionless
non-interacting particles, hereafter referred to as dark matter
(DM). The resulting collapsed objects are dubbed here as {\it
halos}. This seemingly simple problem does not easily yield itself to
an analytical understanding --- the problem addressed here in a
series of numerical `experiments'.

Analytical and numerical studies of the collapse and virialization of
structures in an expanding universe date to the early times of modern
cosmogony.  The only rigorous and exact analytical solution relevant
to the problem is that of a single scale free spherical density
perturbation in a Friedmann universe, the so-called secondary infall
model \citep{gunn77, fg84, bert85} and its application to cosmological
models \citep[][etc.]{hs85}. The shortcoming of
the analytical approach prompted the study of the formation of DM
halos by means of N-body simulations \citep[][etc.]{white76}. 
The structure of DM halos inferred from a
variety of cosmological models and power spectra of the primordial
perturbation field was found to be well approximated by a
spherically averaged density profile \citep[][hereafter NFW]{nfw96,nfw97},
\begin{equation}
\label{eq:nfw}
\rho(r) = \frac{4 \rho_{\rm s}} {(r/R_{\rm s}) 
     (1 + r/R_{\rm s})^2} \,,
\end{equation}
where $\rho_{\rm s}$ is the density at the scale radius \rs. The NFW profile
constitutes a universal fit to the structure of DM halos over many
orders of magnitudes in mass and at different redshifts. It
has been confirmed by numerous N-body simulations, although some
modifications have been suggested \citep[most notably][]{moore98}.
The origin of the NFW profile has been studied in the framework of the 
secondary infall model \citep[\eg][]{ns99, lh00, nusser01, ascasibar04}
and the merger scenario \citep{ez05}. While analytical models necessarily 
invoke spherical symmetry, simulations emphasize the background asymmetry. 
This has led us to embark on a series of numerical experiments carefully 
designed to shed light on the problem.

In this {\it Letter} we address the role of the merging history in
shaping up the density profile. It has been determined that the
evolution of DM halos proceeds in two phases, of rapid and slow
accretion \citep[and refs. therein]{w02,z03, zhao03,ss05}. The general
understanding that has followed is that an NFW structure is quickly
established after the rapid phase and is preserved during the slow
accretion. Thus it follows that the emergence and evolution of the NFW
profile might depend primarily on the mergers epoch \citep[\eg][]{ez05}. 
This motivates our first study of the formation of DM halos by
using constrained simulations. We design a set of numerical
experiments in which a given halo of mass $~10^{12}\hmsun$ (where $h$
is the Hubble's constant in units of $100 \kmsmpc$) is constrained to
follow different merging histories. We then study the different
constrained halos and compare the evolution of their density
profiles.

The ability to design initial conditions for \nbody simulations by
constrained realizations of Gaussian fields is the {\it key} to the
`experimental' approach used here. This is done by following the
\citet{hofrib91} algorithm. The design of an experiment starts with the
expression of whatever constraints one would like to impose into
linear constraints on the primordial perturbation field, which is
assumed to be Gaussian within the inflation paradigm. The constrained
realization is used to set up the initial conditions by means of the
\cite{zeldovich} approximation. The resulting simulation is referred 
here as a constrained simulation. 

The linear constraints are described in \S \ref{sec:cs}, the models are 
presented in \S \ref{sec:model} and their analysis 
in \S \ref{sec:results}. The cosmological implications are 
discussed in \S \ref{sec:disc}.

\section{Constrained Simulations}
\label{sec:cs}

We used the updated version of the FTM-4.4 hybrid code (Heller \& Shlosman 1994;
Heller 1995), with $N\sim 2.1\times 10^6$.
The gravitational forces are computed using the routine {\tt falcON} (Dehnen
2002), which is about ten times faster than optimally coded Barnes \& Hut
(1986) tree code. The gravitational softening is $\epsilon = 500$~pc.  
The addressed issue of the collapse of an individual halo in an expanding 
Friedmann universe has 
led us to use vacuum boundary conditions and to perform the simulations with 
physical coordinates. In these coordinates the cosmological constant, or its
generalization as dark energy, should be introduced by an explicit
term in the acceleration equation. The FTM code in its present form
does not contain a cosmological constant term. This has led us to
assume the open CDM (OCDM) model with $\Omega_0=0.3$, $h=0.7$ and
$\sigma_8=0.9$ (where $\Omega_0$ is the current cosmological matter
density parameter and $\sigma_8$ is the variance of the density field
convolved with a top-hat window of radius $8\hmpc$ used to normalize
the power spectrum).  
This is very close to the `concordance' $\Lambda$CDM model in dynamical
properties. We are interested here in the dynamical evolution of the
density profile and its dependence on the merging history and
therefore the results obtained here are valid also in a $\Lambda$CDM
cosmology. The code was tested in the cosmological context using the
Santa Barbara Cluster model (Frenk et al. 1999).

A series of linear constraints on the initial density field are used
to design the numerical experiments. All the constraints are of the
same form, namely the value of the initial density field at different
locations, and evaluated with different smoothing kernels.  We used
Gaussian kernels for the smoothing procedure, where the width of the
kernel is fixed so as to encompass a mass $M$ (the mass scale on which
a constraint is imposed). The set of mass scales and the location at
which the constraints are imposed define the numerical experiment.

Assuming a cosmological model and power spectrum of the primordial
perturbation field, a random realization of the field is constructed
from which a constrained realization is generated using the
\citet{hofrib91} algorithm. Many different
realizations of the same experiment can be performed (Romano-D\'{\i}az 
\etal\ in prep.).

\section{Models}
\label{sec:model}

A set of five different models, \ie\ experiments, is designed here
to probe different merging histories of a given
$10^{12}\hmsun$ halo in an OCDM cosmology. This halo is then
constrained to have different substructure on different mass scales
and locations designed to collapse at different times. The spherical
top-hat model is used here to set the numerical value of the
constraints. The  model
provides the collapse time of substructures as a function of the
initial density. This is used only as a general rough guide as the
various substructures are neither spherical nor isolated. Furthermore,
the few constraints used here do not fully control the
experiments. The nonlinear dynamics can in principle affect the
evolution in a way not fully anticipated from the initial
conditions. Even more important is the role of the random component of
the constrained realizations \citep{hofrib91}. Thus depending on the
nature of the constraints and the power spectrum assumed, the random
component can provide other significant substructures at different
locations and mass scales. This can be handled by adding more
constraints and varying their numerical values. 

Our models are designed as follows: Model A (our
benchmark model) is based on two constraints. One is that of a
$10^{12}\hmsun$ halo at the origin designed to collapse at
$z_{\rm coll}=2.1$. This halo is embedded in a region (2nd constraint)
corresponding to a mass of $10^{13}\hmsun$ in which the over-density
is zero --- a region corresponding to an unperturbed Friedmann
model. This is a scale larger by about a factor of three (in mass)
than the computational sphere and therefore the constraint cannot be
exactly fulfilled, yet it constrains the large scale modes of the
realizations to obey it. These two constraints are imposed on all
other models.  Model B adds two substructures of mass $5\times
10^{11}\hmsun$ within the $10^{12}\hmsun$ halo, designed to collapse
by $z_{\rm coll}=3.7$.  Model C further splits each ones of the $5\times
10^{11}\hmsun$ halos into two $2.5\times 10^{11}\hmsun$ substructures
($z_{\rm coll}=5.7$). Thus, the benchmark halo is design to follow two
major mergers events on its way to virialization. Model D takes the
Model A and imposes six different small substructures of mass
$10^{11}\hmsun$ scattered within the big halo, designed to collapse at
about $z_{\rm coll}=7.0$.  Model E attempts to simulate a more monolithic
collapse in which a nested set of constraints, located at the origin,
is set on a range of mass scales down to $M=10^{10}\hmsun $
($z_{\rm coll}= 6.9$).  All models have been constructed with the same
seed of the random field.
All the density constraints constitute $(2.5-
3.5)\sigma$ perturbations (where $\sigma^2$ is the variance of the
appropriately smoothed field), and were imposed on a cubic grid of 128
grid-cells per dimension (Romano-D\'{\i}az \etal, in prep.). We evolve 
the linear initial density field from $z=120$ until
$z=0$ by means of the FTM code.  Since we want to follow as close as
possible the merging history of our models, we have sampled the
system's dynamical evolution with 165 time outputs spaced
logarithmically in the expansion parameter $a$. Each halo is resolved 
at $z=0$ with around
$1.2\times 10^6$ particles within the virial radius.  


\section{Results}
\label{sec:results}
 
All models differ substantially at early epochs, with a 
subsequent convergence in some of their properties but not in others.
They lead to the formation of a single object of mass $\sim 10^{12}\hmsun$
via mergers with the surrounding substructure and with a slow accretion.
To analyze the clumpy substructure we define the DM halo(s) as having 
the mean density equal to some critical value $\Delta_{\rm c}$ times the critical 
density of the universe, where $\Delta_{\rm c}$ depends on the redshift and the 
cosmological model.  The  top-hat model is used to calculate
$\Delta_{\rm c}(z)$ and the  density is calculated within a
virial  radius ($R_{\rm vir}$). The halos are identified
initially by the HOP algorithm \citep{hop} and approximated 
by means of a radius $R_{\rm vir}$. Comparison of the HOP 
halos with those of the standard FoF halo finder \citep{defw85} shows good 
agreement. Given a group catalog, a merger tree is constructed for each model 
using all the  snapshots.

\begin{figure*}[!!!!!!!!!!!!!!!!!!!!!!!t]
\epsscale{0.95}
\plotone{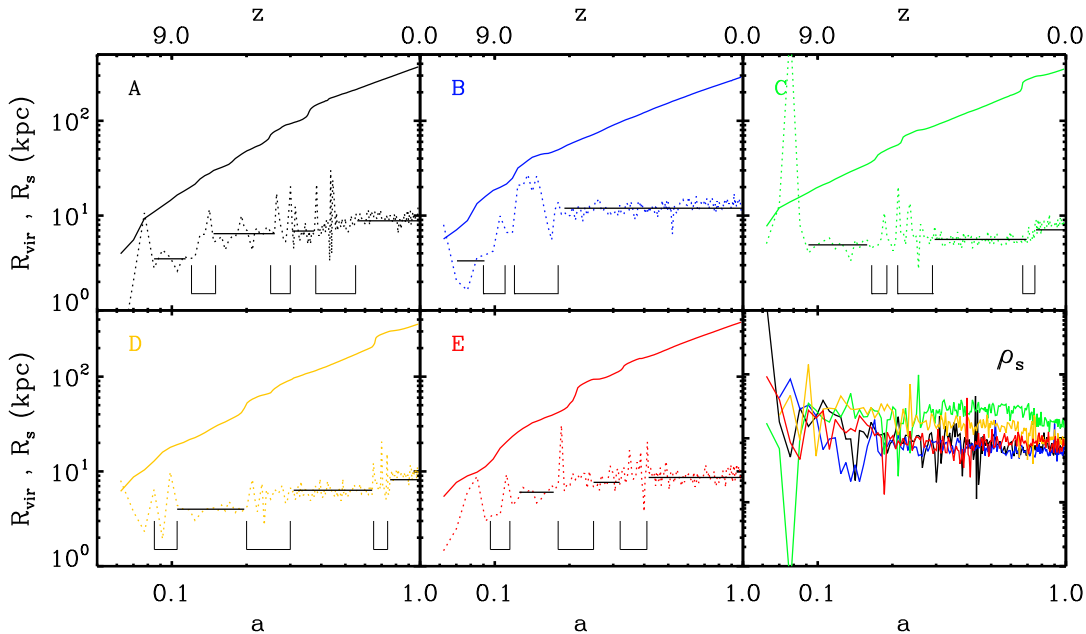}
\caption{Virial and scale radii behavior (continuous and dotted lines 
respectively) as function of $a$ for models A, B, C, D and E. The discontinuous 
growths in \rs\ and \rvir\ match the violent phases that each halo passes through. 
The horizontal bars represent the mean value of \rs within the quiescent phases. 
The square brackets delineate the violent phases. The bottom right panel
shows the evolution of $\rho_{\rm s}$ with colors corresponding to the models
in other panels.
}
\label{fig:rad}
\end{figure*}

The evolution toward the final halo is studied by tracking back in
time the main branch that leads to this halo. The general picture which
emerges is that of a halo undergoing phases of {\it slow} and ordered evolution
intermitted by episodes of rapid mass growth {\it via} collapse and
{\it major mergers}, in agreement with previous simulations 
\citep[\eg][]{w02,zhao03}. These are referred to as the quiescent and violent 
phases.

The structure of a halo is studied by fitting it to an NFW profile
(Eq.~\ref{eq:nfw}) and following the cosmological evolution of the NFW
parameters, namely $R_{\rm s}, R_{\rm vir}$.
Note, that within a given
cosmology and per a given halo of a given mass, only \rs\ is a free
parameter to be fitted. The fitting algorithm used here is based on
logarithmic binning of the halo into spherical shells, and estimating
\rs\ by minimal $\chi^2$, where the residual in a given shell is
normalized by its own density. The fitting is performed within  
$\min(0.6R_{\rm vir}, 0.5d_{\rm H})$, where $d_{\rm H}$ is the distance 
to the nearest massive halo. Although the spherical symmetry of a
NFW model ignores some of the dynamical properties of a halo, we use it 
as the first approximation to the halo structure. 
The NFW profile is found to be a very good fit to the spherically-symmetric 
density in the quasi-static phases. During the violent phases, such as mergers 
of two almost equal mass halos or collapse of a few substructures to 
form a single halo, the halos are out of equilibrium and the NFW fit is
only approximate. 

The cosmological evolution of \rs\ and \rvir\ of the main halo for all the 
models is presented in Fig.~1. 
The \rvir trajectories (\ie
the accretion trajectories) show a regular behavior and a 
linear growth, separated by sudden increases, all of which are 
associated with the violent phases. Most strikingly, \rs\ remains 
{\it constant} in the quiescent phases and grows {\it discontinuously} in the 
violent ones. The \rs\ is subject to a $\sim 5\%-10\%$ 
jitter in the quiescent phases which grows stronger during the
violent phases when the halos get out of a dynamical equilibrium.
The scale density $\rho_{\rm s}$ shows a similar behavior, but in the
opposite sense --- remaining constant in the quasi-static phases and 
decreasing abruptly in the violent ones. We note that all models, except 
B, converge to the same value of \rs (within the jitter) at the present epoch 
of $a\sim 0.8-1$, in spite of the different tracks leading to it.  
In Model B, the two major halos have already turned around but have yet to merge.
Even before this last merger, \rs\ is larger than in other models and is
expected to grow further.

\begin{figure}[!t]
\epsscale{1.1}
\plotone{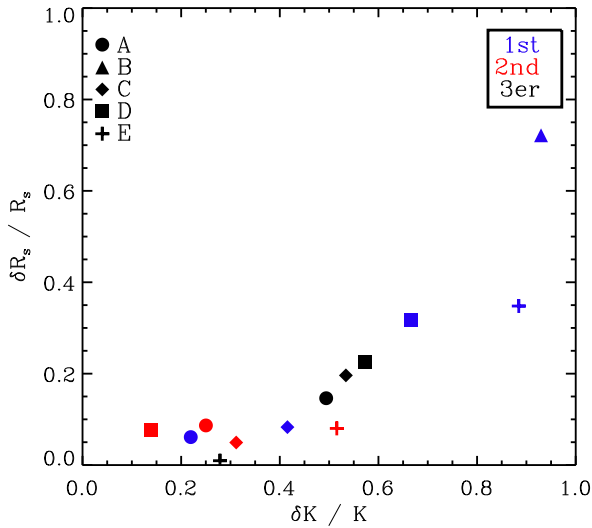}
\caption{The fractional variations in the internal kinetic energy $\delta K/K$ vs
those in the scaling radius $\delta$\rs/\rs\ before
and after the violent phases. The colors correspond to different generations of
major mergers.}
\label{fig:dkei}
\end{figure}

The present simulations show that there is no clear relation between 
$\rho_{\rm s}$ and the formation time of the halo, as given by
the time of the last violent event. This contrasts with previous claims 
about such a correlation.
The order of models given by the formation time of their halos is B, E, A, 
D, and C. All the models have similar $\rho_{\rm s}$ except for Model C 
which has a value twice as large (Fig.~1).

The evolution of the dynamical state of the main halos has been analyzed
in terms of the internal kinetic energy ($K$) within \rs\ and
$R_{\rm vir}$.  
Any perturbation in the halo's internal state (violent
mass aggregation, random energy acquisition, etc.) will be reflected in
its $K$ behavior. We find that $K$ within \rs\ behaves similarly to \rs\ --- 
remaining constant during the quasi-static stages
(within the jitter) and growing discontinuously at
the violent phases. This  implies  a
possible direct relation between the internal kinetic energy and $R_{\rm s}$. 
This has been tested by comparing the relative changes of $K$ and 
\rs\ (Figure~2), which shows that the larger is the  change 
in $K$ the more \rs\ increases. (Further analysis of this relation 
is to be presented elsewhere.)
The internal kinetic energy computed within \rvir\
shows a similar behavior 
but grows very slowly during the quasi-static phases.


\section{Discussion}
\label{sec:disc}

The halo growth can be divided into the violent and quiescent
phases analyzed in our superior time sampled
$N$-body simulations using constrainted realizations. This allowed us to 
show conclusively  
the details which have been hinted about in the literature so far. In this
{\it Letter} we focus on the evolution of \rs\ and $R_{\rm vir}$
leaving more comprehensive analysis for elsewhere (Romano-Diaz \etal, in prep.).
First, we find that the NFW scale \rs\ stays constant during the quiescent
phase and changes abruptly during the violent one. In contrast, $R_{\rm vir}$
is growing linearly in the quiescent and abruptly during the violent
phases. Second, $\rho_{\rm s}$ stays unchanged during the quiescent
phase and drops abruptly during the violent phase. 
Third, the value of \rs\ reflects the violent merging history of the halo,
and depends on the number of violent events and their fractional magnitudes,
independent of the time and order of these events. The corollary is
that $\rho_{\rm s}$ does not reflect the formation time of the halo.
Fourth, the fractional change in \rs\ is a nonlinear function of the
fractional absorbed kinetic energy within \rs\ in a violent event.
We note, that the accretion trajectories in all models converge to the same
value. This is a reflection of the large-scale structure shared by 
all the models and imposed by the constrained initial conditions.

The advantage of the constrained realizations lies in the unique ability
to generate a series of models with one or more parameters varied
in a controlled manner, while all others are held fixed. It allows us to
cleanly separate the cause-and-effect  relationship between the initial 
parameters and the outcome of the dynamical evolution. This complements
the prevailing method of large-scale cosmological simulations in which issues of
structure and evolution are addressed statistically. The scaling relations 
found in such a statistical analysis are not necessarily applicable 
to an individual halo.
Here we focus on the role of the merging history in the halo evolution,
by imposing density constraints based on the top-hat model. We find that
the actual history of a halo does not always follow closely the expectations 
based on the simple top-hat model. Inspite of this the different models provide 
us with a good laboratory for experimenting with the role of the merging history
in shaping the structure of DM halos.   
 
The analogy between the halo evolution and thermodynamical processes has not
escaped our attention. Equating the quiescent phases with adiabatic
processes and the violent with non-adiabatic ones leads one 
to associate the behavior of \rs\ with that of the entropy. In this 
terminology the entropy remains constant in the quiescent phase and grows
discontinuously in the violent phase (see Fig.~2). 
Also,  the accretion 
trajectories play the role of  adiabats and the system jumps from one 
adiabat to the other by a violent event, not unlike a shock wave. The question
whether this is just a simple analogy or provides a deeper
understanding will be addressed elsewhere.

The results obtained here pertain to the one halo studied
in the framework of an OCDM cosmology. Yet, the conclusions reached
from the set of experiments presented here are relevant to the
understanding of halo formation in the general CDM cosmologies and in
particular to the `benchmark' $\Lambda$CDM cosmology.


\acknowledgments
This research has been partially supported by ISF-143/02 and the 
Sheinborn Foundation (YH), by NSF/AST 00-98351 and NASA/NAG 5-13275 (AF),
by NASA/LTSA 5-13063, NASA/ATP NAG5-10823, HST/AR-10284 (IS),
and by NSF/AST 02-06251 (CH \& IS). ERD
has been partially supported by the Golda Meier fellowship at the HU.
We acknowledge fruitful discussions with A. Klypin \& R.v.d.
Weygaert.


\end{document}